%% file: simulator17techreport.tex
\documentclass[sigchi]{acmart}

\usepackage{booktabs} 
\usepackage{tabularx, booktabs} 
\usepackage{color}
\usepackage{amsmath}
\usepackage{graphicx}
\usepackage{caption}
\usepackage{subcaption}
\usepackage[inline]{enumitem}
\usepackage{balance}

\newcommand\tab[1][1cm]{\hspace*{#1}}
\newcommand{\ra}[1]{\renewcommand{\arraystretch}{#1}}
\renewcommand\footnotetextcopyrightpermission[1]{} 
\makeatletter
\renewcommand\@formatdoi[1]{\ignorespaces}
\makeatother

\begin{document}
\title{ThermalSim: A Thermal Simulator for Error Analysis}

\author{Milan Jain}
\authornote{The author did the work during an internship at University of Waterloo.}
\affiliation{%
  \institution{IIIT-Delhi, India}
}
\email{milanj@iiitd.ac.in}

\renewcommand{\shortauthors}{M. Jain et al.}

\begin{abstract}
Researchers have extensively explored predictive control strategies for controlling  heating, ventilation, and air conditioning (HVAC) units in commercial buildings. 
Predictive control strategies, however, critically rely
on weather and occupancy forecasts. Existing state-of-the-art building simulators are incapable of analysing the influence of prediction errors (in weather and occupancy) on HVAC energy consumption and occupant comfort. In this paper, we introduce ThermalSim, a building simulator that can quantify the effect of prediction errors on the HVAC operations. ThermalSim has been implemented in C/C++ and MATLAB. We describe its design, use, and input format.
\end{abstract}

%
%




\maketitle

\input{simulator17techreport_body}

\bibliographystyle{ACM-Reference-Format}
\bibliography{sigproc} 

\end{document}

%% file: simulator17techreport_body.tex
\section*{Introduction}
Buildings consume one-third of total energy across the world, with heating, ventilation, and air-conditioning (HVAC) being the major contributor~\cite{jain2016data}. Recent studies~\cite{scott2011preheat, gao2013optimal, jain2017portable+} show that predictive control strategies (for HVAC) can significantly reduce the energy footprint without compromising the user comfort. However, predictive control strategies primarily rely on the accuracy of weather and occupancy forecast~\cite{jain2016non}. Therefore, it becomes critical for the researchers to either conduct the studies in real-world or use building simulators to analyse the influence of prediction errors on the effectiveness of their proposed approach. 

In this work, we present ThermalSim - a framework for researchers to simulate the thermal behaviour of a commercial building for error-analysis. The following features differentiate ThermalSim from existing simulators, 
such as EnergyPlus~\cite{crawley2000energyplus} and GridLAB-D~\cite{gridlabd}:
\begin{itemize}
	\item Lightweight: ThermalSim is implemented in C/C++ (along with MATLAB) while using header based libraries to ensure that it can be executed even on low-cost Single Board Computers (SBC) such as Raspberry Pi. 
	\item Compatible: ThermalSim can be run on Linux, Windows, or Mac.
	\item Easy-to-Use: ThermalSim takes human readable key-value pairs as an input. Users can run any number of instances in parallel, giving different files as input. 
	\item Easy-to-Customize: Users can add thermal models and control algorithms by writing simple functions in C/C++ and MATLAB. 
\end{itemize}

Though we designed ThermalSim primarily for modelling thermal behaviour of a building, it is possible to extend the framework to model other applications. For example, ThermalSim can be used to simulate the heat dissipation across battery cells.

Next, we discuss the architecture of ThermalSim. 

\begin{figure*}
	\begin{subfigure}[t]{0.4\textwidth}
		\includegraphics[width=\textwidth]{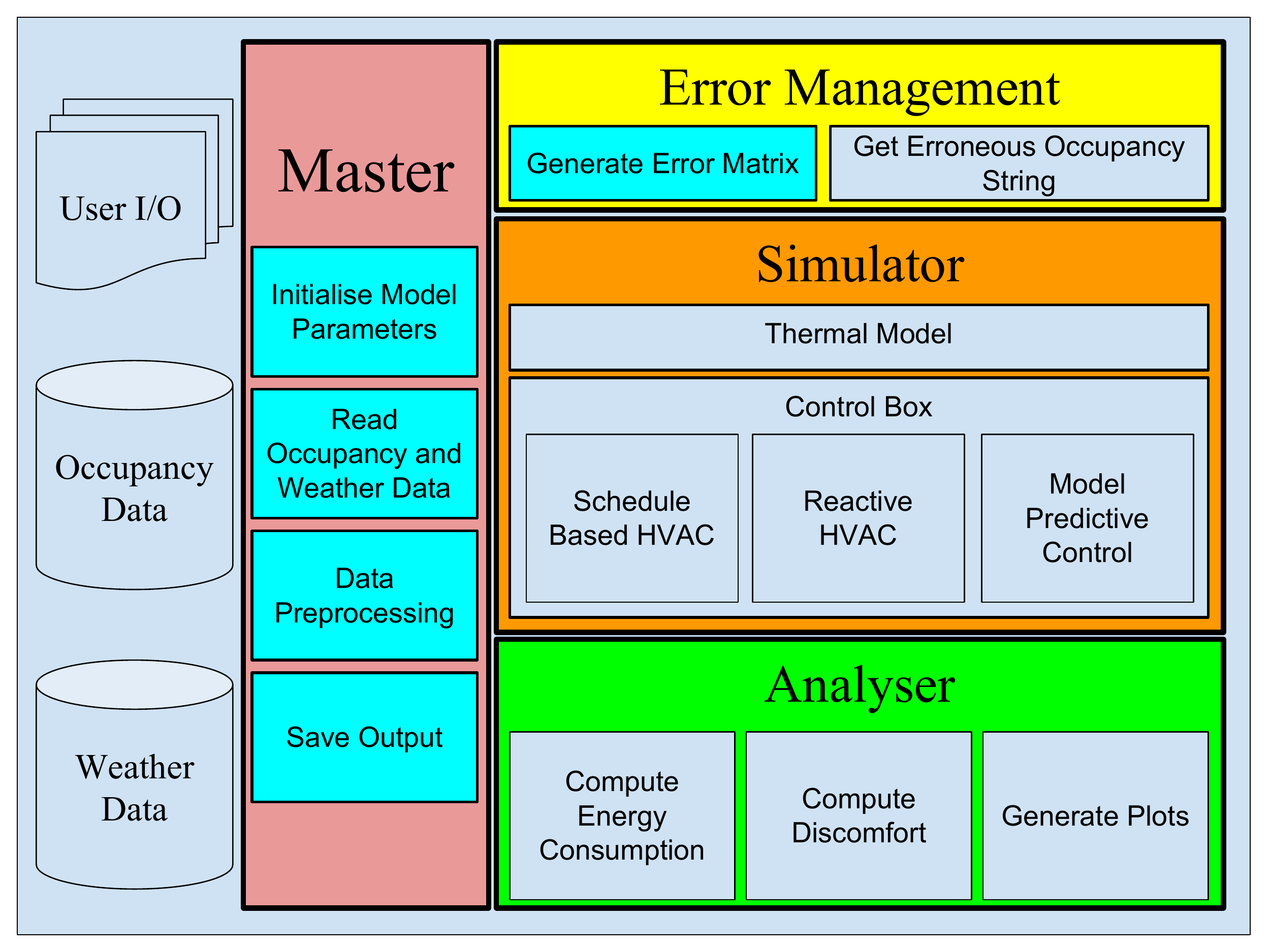}
		\caption{Architecture}
		\label{fig:architecture}
	\end{subfigure}
	\begin{subfigure}[t]{0.59\textwidth}
		\includegraphics[width=\textwidth]{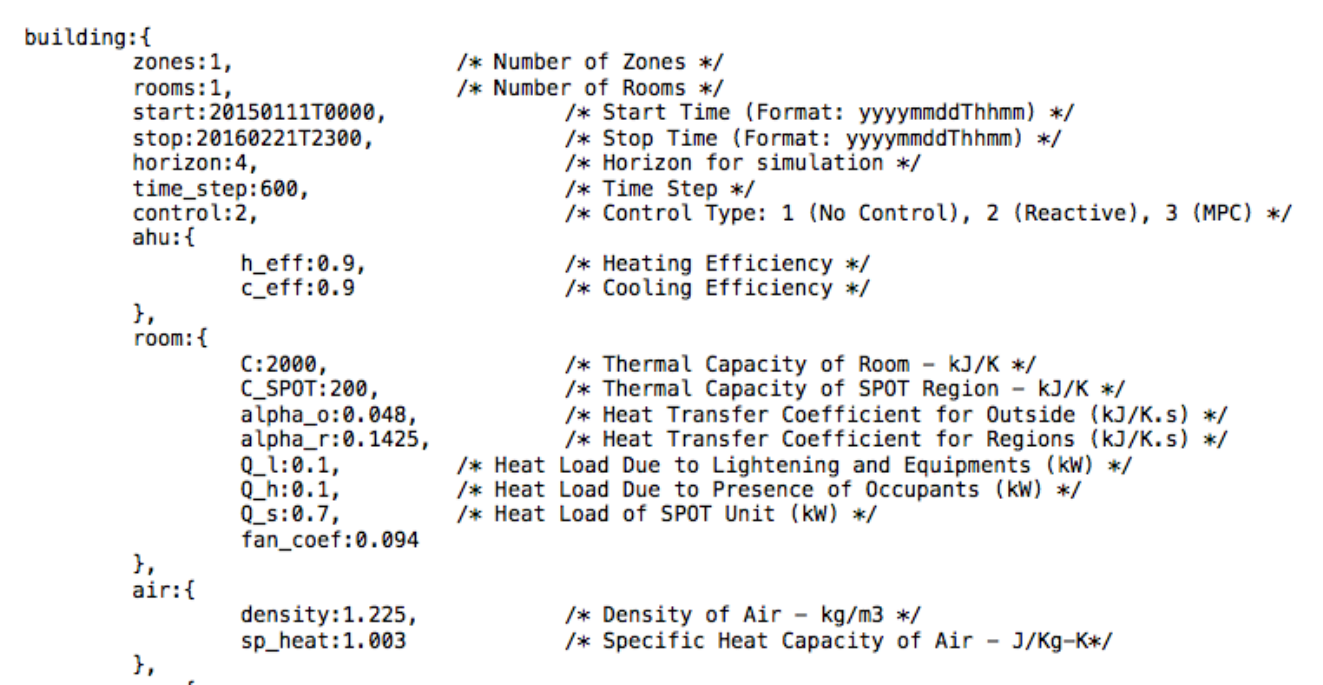}
		\caption{Input format}
		\label{fig:input}
	\end{subfigure}
	\caption{ThermalSim - Building Simulation Framework}	
	\label{fig:simulator}
\end{figure*}

\section*{System Architecture}
Figure~\ref{fig:architecture} depicts the detailed design of ThermalSim. It consists of four major modules - 
\begin{enumerate}
	\item \emph{Master} - handles data input/output and preprocessing,
	\item \emph{Error Management} -  injects \textit{realistic} errors in occupancy data,
	\item \emph{Simulator} - simulates temperature evolution for a given thermal model and control logic, and
	\item \emph{Analyser} - computes the energy consumption and discomfort based on the simulation.
\end{enumerate}
To estimate control parameters for the HVAC (in the \emph{simulator} module), we also incorporated AMPL~\cite{AMPL} --an algebraic modelling language for mathematical programming --into ThermalSim. 

\subsection{Master}
ThermalSim takes as input historical weather and occupancy data in CSV (Comma Separated Values) format and a user-generated 
description of building and simulation control 
parameters in a separate file in the form of key-value pairs (Figure~\ref{fig:input}). The latter input includes start/stop time for the simulation, parameters for the thermal model, and choice of control strategies, among others. 
Based on the \emph{start} and \emph{stop} time of the simulation (as provided by the user), the \emph{master} module slices the weather and occupancy data for the corresponding duration. Depending on the data requirements, the \emph{master} module preprocesses the sliced data set, prior executing the simulations (filling in missing values, upsampling, downsampling) . Finally, \emph{master} module saves the output of the simulation and the analysis, to various files. 

\subsection{Error Management}
ThermalSim represents occupancy data for each day as a string of consecutive 0's (for unoccupied workspaces) and 1's (for occupied spaces). We call this string an \emph{occupancy string}. We only consider two states of occupancy because a majority of occupancy prediction algorithms use occupancy as a two state variable. The length of a single occupancy string depends upon the sampling rate of the occupancy data. Data sampled every ten minutes will generate an occupancy string of length 144 characters and data sampled every thirty-seconds will result in 2880 character-long string. ThermalSim divides the whole occupancy data set into day-wise occupancy strings to then generate an \emph{error matrix} for a systematic injection of errors in the occupancy data.

\subsubsection{Error Matrix}
The goal of the Error Matrix is to add occupancy
errors in a realistic manner. That is,
the process of adding occupancy errors should result
in a realistic occupancy trace, rather than an unrealistic
one, such as occupancy during the middle of the night.

Specifically, given a dataset with $n$ occupancy strings, an \emph{error matrix} is a symmetric matrix of size $n^2$ with diagonal elements equal to zero. Each matrix
element is the percentage difference between any two occupancy strings, using the Hamming distance metric. This is the number of positions at which there is a mismatch in the given strings~\cite{hamming1950error}. NExt, ThermalSim divides the Hamming distance by the total length of the \emph{occupancy string} to compute the error percentage.

To illustrate, suppose a user wants to analyse different control strategies given a 10\% error in occupancy forecast. The \emph{error management} module will look for an occupancy string (in the database) which is closest to the day of analysis. We term the selected occupancy string as the \emph{reference} string. The module will then look into the \emph{error matrix} to find all those strings that have 10\% error when compared with the \emph{reference string} and randomly selects one of them. We term the selected string as the {erroneous} string. If the day (\emph{reference} string) is 30\% occupied, then total occupancy in the \emph{erroneous} string may fall in the range of 20\%-40\%. The \emph{error management} module finally returns the \emph{reference} and \emph{erroneous} string for further processing. 

\subsection{Simulator}
The \emph{simulator} module takes input from the \emph{master} and \emph{error management} modules to simulate the thermal behaviour of a given space. It comprises of two major blocks - 
\begin{enumerate*}
	\item thermal model - depicts the  thermodynamics of the space, and
	\item control module - computes control variables. 
\end{enumerate*} 

\subsubsection{Thermal Model}
The room temperature depends upon the external temperature, occupants, and various heating or cooling loads present in the room. A thermal model depicts the heat exchange happening within the room. The simulator supports a general HVAC architecture for a building consisting of multiple air handling units (AHUs) where each AHU regulates temperature in various variable air volume (VAV) zones. Each VAV zone provides hot/cold air across all the rooms (and shared spaces) present in that particular zone.

\subsubsection{Control Box}
A building manager uses the control parameters (such as supply air volume, supply air temperature) as a knob to achieve the desired comfort levels in the room. These parameters also guide the total energy consumption (of HVAC) and occupants' comfort. The value of these parameters depend on the control strategy, implemented to control the HVAC.

\subsection{Analyser}
In this section, we discuss the metrics to compute energy consumption and the discomfort.

	\begin{equation}
		E = \sum_{t=0}^{n_t} P{(t)} \times \frac{\tau}{3600}
		\label{eq:energy}
	\end{equation}
	
\begin{enumerate}
	\item \textbf{Energy Consumption:} This is the energy consumed by a building in a day (Equation~\ref{eq:energy}),
    where $P(t)$ denotes the power consumed by HVAC (and other heating/cooling devices), $\tau$ denotes the sampling rate, and $n_t$ denotes the total number of data samples in a day. 

 	\item \textbf{Occupants' Discomfort:} To compute discomfort, ThermalSim leverages Predicted Mean Vote (PMV)~\cite{ASHRAE} that refers to a thermal scale to decide if the occupants will be satisfied in the given thermal conditions or not (Equation~\ref{eq:pmv}). 
	\begin{equation}
	\begin{split}
		P^{ij}{(t)} &= P1 \times T_{oc}^{ij}{(t)} - P2 \times v_{a}^{ij}{(t)}\\ &+ P3 \times v_{a}^{ij}{(t)} \times v_{a}^{ij}{(t)} - P4 \\{}
	\end{split}
	\label{eq:pmv}
 	\end{equation}
  	
 	At a given time instance $t$, If PMV ($P^{ij}(t)$) lies with the comfort requirements ($[P_{ll}-P_{ul}]$) of an individual then room $j$ or VAV zone $i$ is marked as comfortable else uncomfortable (Equation~\ref{eq:dc}). 

	\begin{equation}
	\begin{split}
		D^{ij}{(t)} = max(0, P_{ll} - P^{ij}(t), P^{ij}(t) - P_{ul}) \\{}
	\end{split}
	\label{eq:dc}
 	\end{equation}
 	
 	$D_{\%}^{ij}$ denotes the total percentage of time instances (in a day) when the user experienced discomfort whenever the room was occupied (Equation~\ref{eq:dcp}). 
 	
 	\begin{equation}
	\begin{split}
		D_{\%}^{ij} = \frac{\sum_{t=0}^{n_t}[{D^{ij}{(t)} \ne 0}]}{\sum_{t=0}^{n_t}[{O^{ij}{(t)} = 1}]} \\{}
	\label{eq:dcp}
	\end{split}
 	\end{equation}
 \end{enumerate}

Table~\ref{table:aparams} summaries the terms used in the analysis. The analyzer utilizes the specified metrics to compare different control strategies, as defined in its controller. Besides, it also provide various scripts to generate comparative plots to facilitate deeper analysis. 

\input{aparams}

\begin{figure}[h!]
	\includegraphics[width=\columnwidth]{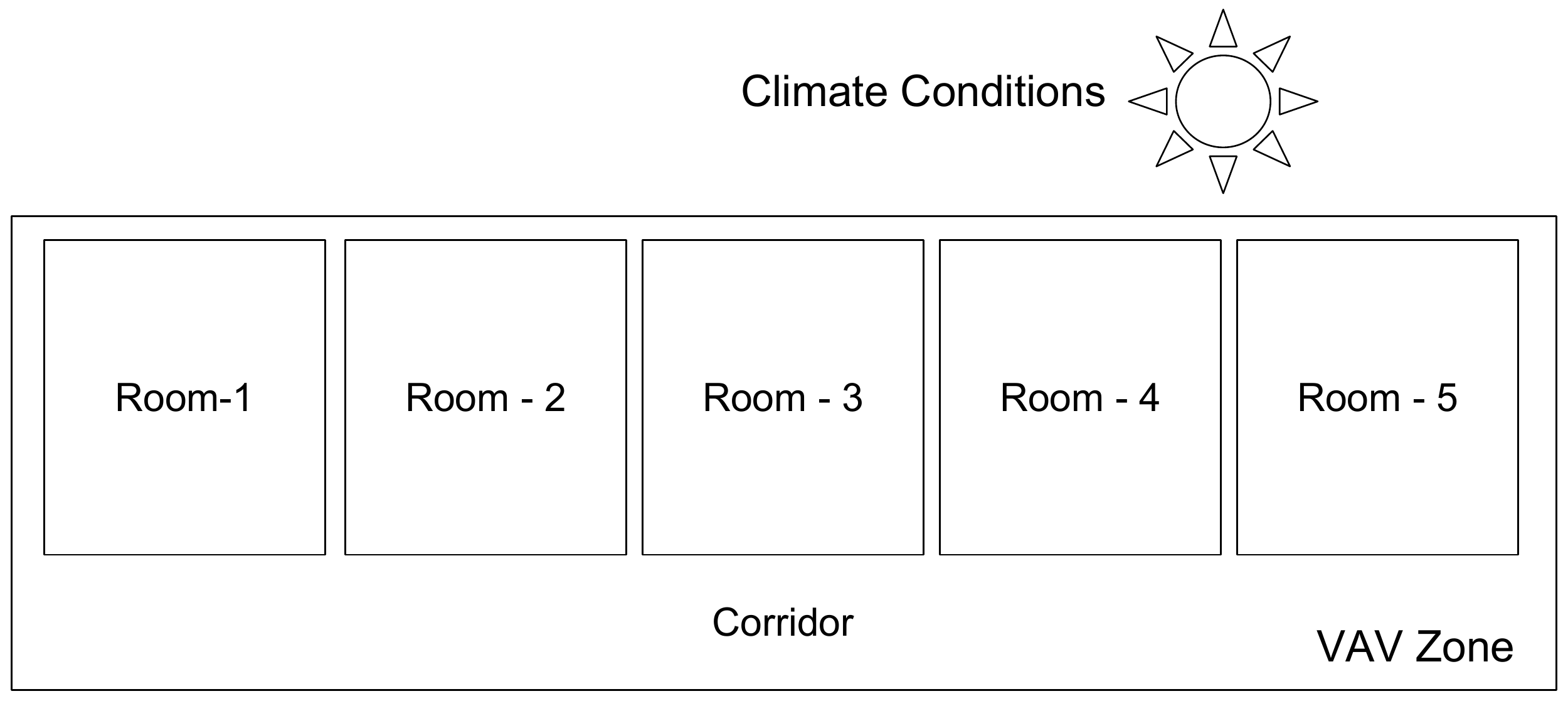}
	\caption{A VAV zone comprising of five rooms}	
	\label{fig:setup}
\end{figure}

\begin{figure*}
	\begin{subfigure}[t]{0.63\textwidth}
		\includegraphics[width=\textwidth]{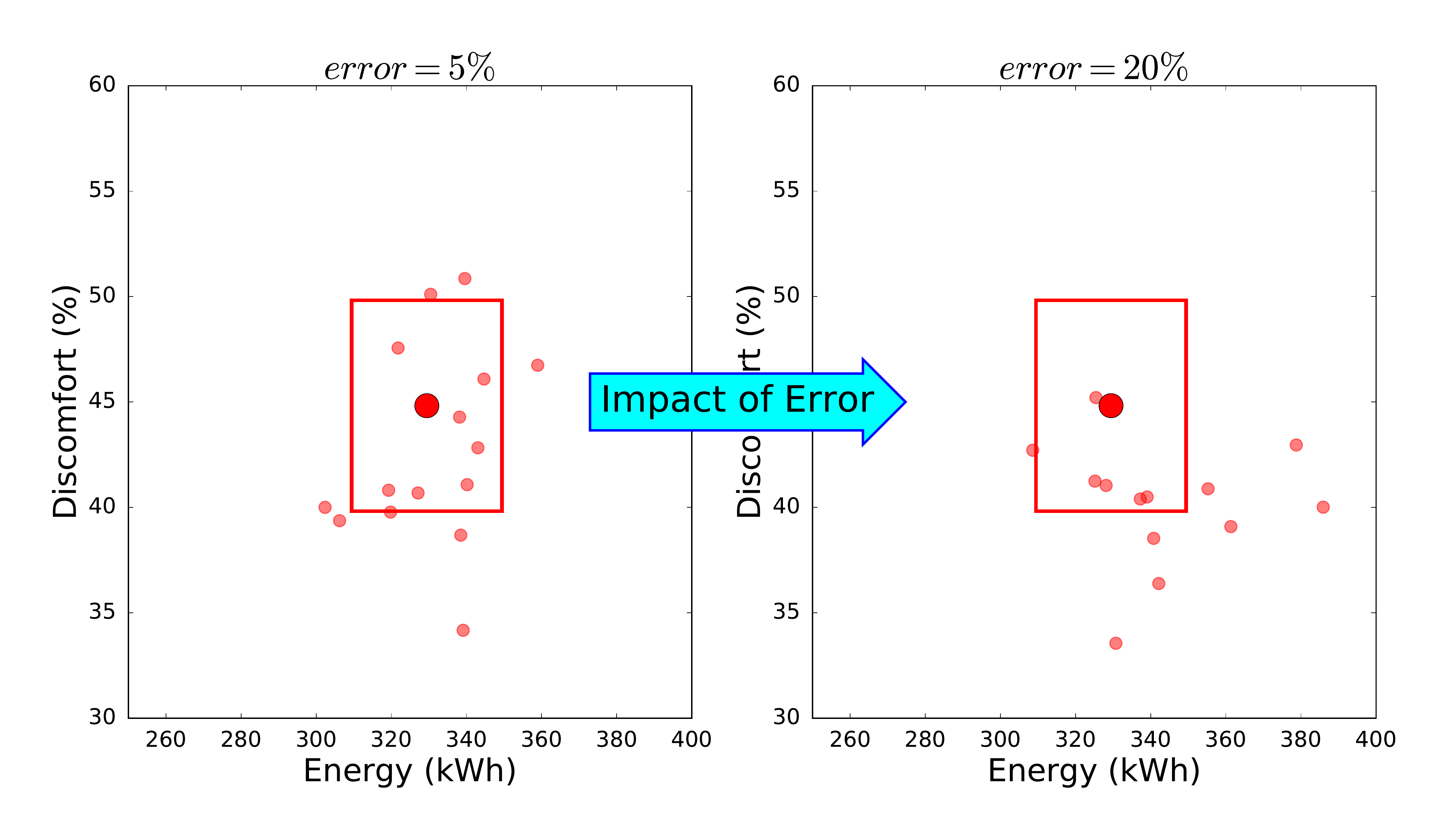}
		\caption{As we increased the forecast errors, we observed significant deviations from the ideal behaviour.}
		\label{fig:illustration}
	\end{subfigure}
	\begin{subfigure}[t]{0.36\textwidth}
		\includegraphics[width=\textwidth]{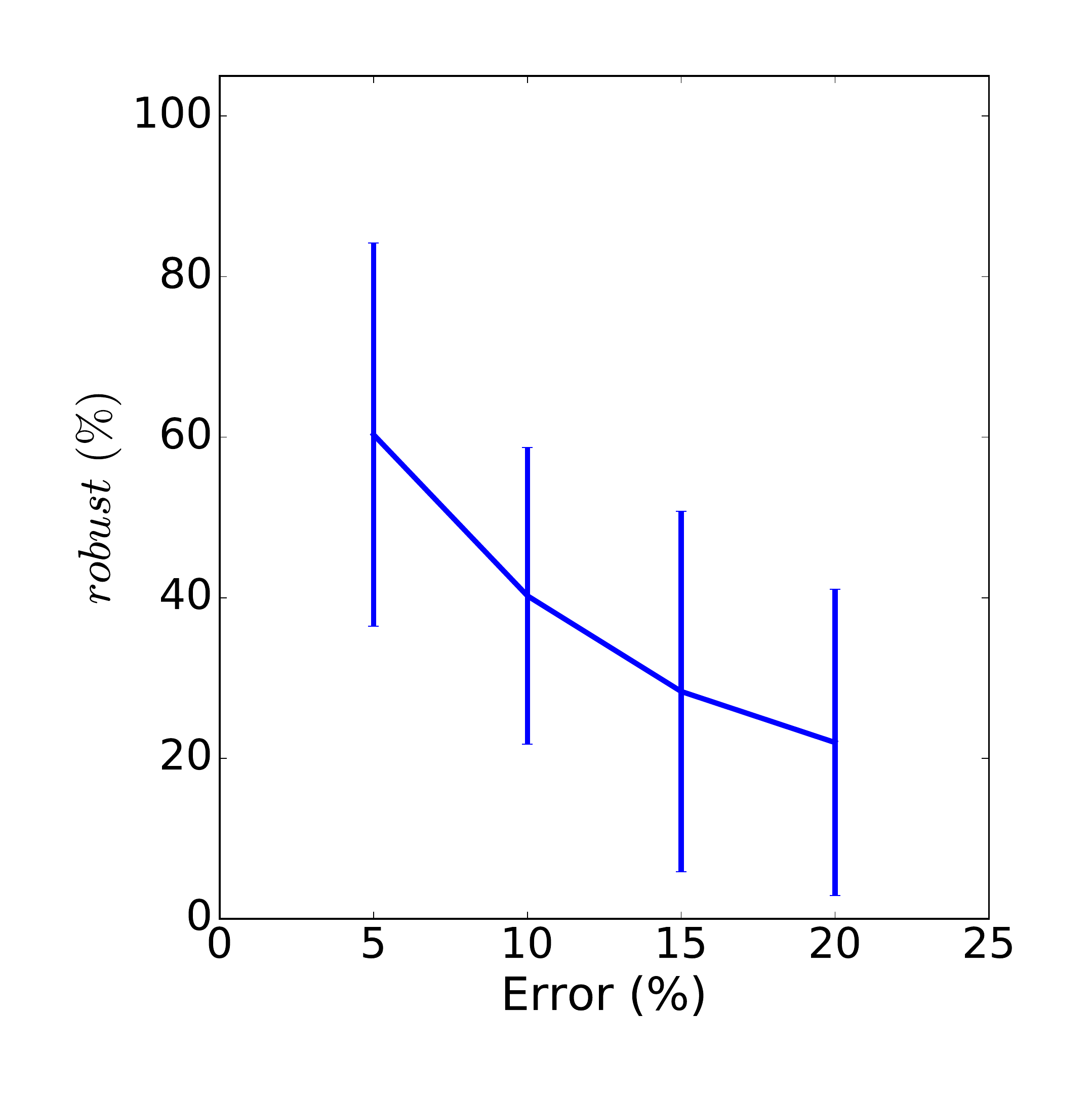}
		\caption{Increase in prediction error ($5\% \rightarrow 20\%$) makes the MPC unreliable.}
		\label{fig:err_wise_summers}
	\end{subfigure}
	\caption{The prediction errors makes MPC a non-deterministic control strategy. The stochastic behavior of prediction errors leads to an instability in the HVAC operations and makes it hard to estimate the energy consumption and user discomfort for a particular instance of control variables.}	
	\label{fig:results}
\end{figure*}

\section*{Case Study}
In this section, we demonstrate ThermalSim's capabilities by evaluating an implementation of model predictive control strategy for different levels of
occupancy error. 

We consider a single zone (of a building) comprising of five rooms surrounded by walls on the three sides and exposed to weather conditions from the fourth side (Figure~\ref{fig:setup}). We assume that there is an AHU unit in the building and a dedicated VAV unit for each zone within the building. Though this building is hypothetical, it is a common architectural pattern, for example, used while designing faculty offices in Universities where offices 
are in series separated by a thick brick wall. In model predictive control (MPC), the optimization is performed over a time horizon, as opposed to just at one point of time, which requires predictions of occupancy and weather for future time steps. 
We implement MPC with two time scales~\cite{kalaimani2016multiple} where HVAC supply air temperature is changed every 1 hour and supply air volume every 10 minutes. These time-scales are determined by the physical limitation of the system. 

Predictive control strategies depend on the occupancy forecast to run the HVAC, and the errors in occupancy forecast can lead to undesired outcomes such as energy wastage and occupants' discomfort. For a given error percentage, there are many possible erroneous occupancy strings, and
any of the ones selected by the ThermalSim does not necessarily lead to an increase in energy, or discomfort. To illustrate, the large circle in Figure~\ref{fig:illustration} depicts the energy consumption and occupants' discomfort for one day while assuming perfect occupancy prediction. The stochastic behavior of prediction errors leads to instability in HVAC operations and makes it hard to precisely estimate  energy consumption and user comfort. Therefore, for each day and error percentage, ThermalSim simulates fifteen different erroneous occupancy patterns.
This compensates for any bias due to the selection procedure of the erroneous occupancy strings. The energy consumption and occupants' discomfort through these erroneous occupancy strings are depicted by the smaller red circles.

We see that when the prediction error increases from $5\%$ (left) to $20\%$ (right), the spread depicting the erroneous strings widens and starts moving away from the point showing the perfect prediction. However, a building manager would prefer limited deviations; thus, to quantify the impact of prediction errors, Equation~\ref{eq:robust} presents a metric $robust$ which computes the number of instances (out of fifteen) that stays within the desired limits of the building manager. 

\begin{equation}
	robust~(\%) = \frac{\text{\# of instances within limits}}{\text{total \# of instances}} \times 100
	\label{eq:robust}
\end{equation} 

We use $\pm 20~kWh$  and $\pm 5\%$ as the acceptable limits for energy consumption and occupants' discomfort, respectively. The red boxes in Figure~\ref{fig:illustration} shows the acceptable limits of energy and discomfort for MPC. 

In case of MPC, the system decides the control parameters such that the desired room temperature (same for each room) is achieved across all the rooms. Therefore, in case of a sudden change in the occupancy, the MPC has to reset the control parameters which might take considerable time to balance the energy-discomfort tradeoff. Therefore, as the prediction error increases from $5\%$ to $20\%$, the performance of MPC keeps dropping (Figure~\ref{fig:err_wise_summers}).

\balance

\appendix
\section{Input for Simulation}
In this section, we present the input for simulation.\\

building: \{ \\
\tab zones: $1$, \\
\tab rooms: $5$, \\
\tab start: 20150101T0000, \\
\tab stop: 20150126T0000, \\
\tab horizon: $4$, \\
\tab time\_step: $600$, \\
\tab control: $2$, // 1 - No Control, 2 - Reactive, 3 - MPC \\
\tab ahu: \{ \\
\tab \tab heating efficiency: $0.9$, \\
\tab \tab cooling efficiency: $0.9$ \\
\tab \}, \\
\tab room: \{ \\
\tab \tab thermal capacity of room: $2000~kJ/K$, \\
\tab \tab heat transfer coefficient for outside: $0.048~kJ/Ks$, \\
\tab \tab heat load due to equipments: $0.1~kW$, \\
\tab \tab heat load due to occupants: $0.1~kW$, \\
\tab \tab coefficient of fan: $0.094$ \\
\tab \}, \\
\tab air: \{ \\
\tab \tab density: $1.225~kg/m^3$, \\
\tab \tab specific heat: $1.003~J/Kg.K$ \\
\tab \}, \\
\tab pmv: \{ \\
\tab \tab p1: $0.2466$, \\
\tab \tab p2: $1.4075$, \\
\tab \tab p3: $0.581$, \\
\tab \tab p4: $5.4468$ \\
\tab \}, \\
\tab error: \{ \\
\tab \tab occupancy: $5\%$, \\
\tab \tab external temperature: $0\%$ \\
\tab \}, \\
\tab files: \{ \\
\tab \tab weather: \textit{<filename>}, \\
\tab \tab occupancy: \textit{<filename>}, \\
\tab \tab output: \textit{<filename>} \\
\tab \} \\
\} \\

The start and stop time is taken in $yyyymmddThhmm$ format. The values of parameters for room rely on the thermal model implemented for the simulations. The parameters of PMV indicate the coefficients learned by running a regressor over a linear model designed to compute to PMV. We will shortly release the code. 
\begin{acks}
  The author would like to thank Prof. S. Keshav, Prof. C. Rosenberg, and Dr. R. Kalaimani for the regular and critical feedback in designing the framework. 
\end{acks}

%% file: aparams.tex
\begin{table}[t]
\centering
\resizebox{\columnwidth}{!} {%
\ra{1.5}
\begin{tabular}{@{}lm{5cm}rl@{}}\toprule[0.3ex]
\textbf{Symbol}				& \textbf{Description} & \textbf{Default} 		& \textbf{Unit}       \\
\hline
$n_t$  							&   Total number of time instances (in a day) sampled at every $\tau$ seconds
& $-$ & $-$ \\
$P{(t)}$  						&  Total power consumed at time $t$  			& $-$ & $kW$\\
$E$  								&  Total energy consumed in a day  				& $-$ & $kWh$\\
$P^{ij}(t)$  						&  PMV in room $j$ of zone $i$ at time $t$  		& $-$ & $-$\\
$v_{a}^{ij}(t)$  				&  Fan speed in room $j$ of zone $i$ at time $t$  		& $-$ & $m/s$\\
$P_{ll}$  							&  Minimum PMV required to maintain user comfort  		& $-$ & $-$\\
$P_{ul}$  						&  Maximum PMV allowed to maintain user comfort 		& $-$ & $-$\\
$D^{ij}(t)$  						&  Discomfort in room $j$ of zone $i$ at time $t$	 & $-$ & $-$\\
$O^{ij}(t)$  						&  Occupancy in room $j$ of zone $i$ at time $t$  		& $-$ & $-$\\
$D_{\%}^{ij}$  				&  Percentage of time occupant felt uncomfortable (in a day) when the room $j$ of zone $i$ was occupied & $-$ & $\%$\\
\bottomrule[0.3ex]
\end{tabular}%
}
\vspace{1em}
\caption{List of parameters used for the analysis}
\label{table:aparams}
\end{table}